\documentclass[12pt,twoside,a4paper]{article}
\usepackage{graphicx}

\def\gsim{\:\raisebox{-0.5ex}{$\stackrel{\textstyle>}{\sim}$}\:}
\begin{document}
\thispagestyle{empty} 
\title{
\vskip-3cm
{\baselineskip14pt
\centerline{\normalsize DESY 11--158 \hfill ISSN 0418--9833}
\centerline{\normalsize MZ-TH/11--23 \hfill} 
\centerline{\normalsize September 2011 \hfill}} 
\vskip1.5cm
\boldmath
{\bf Inclusive $B$-Meson Production at the LHC}
\\
{\bf in the GM-VFN Scheme}
\unboldmath
%\\
%{\bf and Comparison with CMS Data}
\author{
B.~A.~Kniehl$^1$, 
G.~Kramer$^1$, 
I.~Schienbein$^2$ 
and H.~Spiesberger$^3$
\vspace{2mm} \\
\normalsize{
  $^1$ II. Institut f\"ur Theoretische
  Physik, Universit\"at Hamburg,
}\\ 
\normalsize{
  Luruper Chaussee 149, D-22761 Hamburg, Germany
} \vspace{2mm}\\
\normalsize{
  $^2$ Laboratoire de Physique Subatomique et de Cosmologie,
} \\ 
\normalsize{
  Universit\'e Joseph Fourier Grenoble 1,
}\\
\normalsize{
  CNRS/IN2P3, Institut National Polytechnique de Grenoble,
}\\
\normalsize{
  53 avenue des Martyrs, F-38026 Grenoble, France
} \vspace{2mm}\\
\normalsize{
  $^3$ Institut f\"ur Physik,
  Johannes-Gutenberg-Universit\"at,
}\\ 
\normalsize{
  Staudinger Weg 7, D-55099 Mainz, Germany
} \vspace{2mm} \\
}}

\date{}
\maketitle
\begin{abstract}
\medskip
\noindent
We calculate the next-to-leading-order cross section for the inclusive 
production of $B$ mesons in $pp$ collisions in the 
general-mass variable-flavor-number scheme, 
an approach which takes into account the finite mass of the 
$b$ quarks. We use realistic evolved nonperturbative fragmentation 
functions obtained from fits to $e^+e^-$ data and compare our results 
for the transverse-momentum and rapidity distributions at a 
center-of-mass energy of 7 TeV with recent data from the CMS 
Collaboration at the CERN LHC. We find good agreement, in particular 
at large values of $p_T$.
\end{abstract}

\clearpage

%**********************************************************************

\section{Introduction}

Since the late eighties there has been much interest in the study of 
$B$-meson production in $p\bar{p}$ and $pp$ collisions at hadron 
colliders, both experimentally and theoretically. The first measurements 
were performed more than two decades ago by the UA1 Collaboration 
at the CERN $S\bar{p}pS$ collider \cite{Albajar:1988th} operating at a 
center-of-mass energy of $\sqrt{S}=0.63$ TeV. More recent measurements 
were made by the CDF and D0 Collaborations at the Fermilab Tevatron running at
$\sqrt{S} = 1.8$ TeV 
\cite{Abe:1995dv,Abachi:1994kj} and 1.96 TeV \cite{Acosta:2004yw}. Just 
recently, the CMS Collaboration at the CERN LHC collider 
published first results for inclusive $B^+$- \cite{Khachatryan:2011mk},  
$B^0$- \cite{Chatrchyan:2011pw}, and $B_s$-meson \cite{Chatrchyan:2011vh}
production in $pp$ collisions at $\sqrt{S} = 7$ TeV. $B^+$ mesons 
were reconstructed via their decay $B^+ \to J/\psi K^+$ followed by $J/\psi 
\to \mu^+\mu^-$, whereas $B^0$ mesons were identified through the 
observation of $J/\psi K_s^0$ final states with the subsequent decays 
$J/\psi \to \mu^+\mu^-$ and $K_s^0 \to \pi^+\pi^-$. In the case of 
$B_s$ mesons, the reconstructed final states were generated by the 
decay chain $B_s \rightarrow J/\psi \phi$, $J/\psi \rightarrow \mu^+ 
\mu^-$, and $\phi \rightarrow K^+ K^-$. From all these measurements 
the differential cross sections $d\sigma/dp_T$ and $d\sigma/dy$ as 
well as the integrated cross section for $p_T \geq 5$ GeV (for $B^+$ 
and $B^0$ mesons) or $p_T \geq 8$ GeV (for $B_s$ mesons) were reported.

The general-mass variable-flavor-number (GM-VFN) scheme provides a rigorous
theoretical framework for the description of the inclusive production of single
heavy-flavored hadrons, combining the fixed-flavor-number (FFN) \cite{theory}
and zero-mass variable-flavor-number (ZM-VFN) \cite{BKK} schemes, which are
valid in complementary kinematic regions, in a unified approach that enjoys the
virtues of both schemes and, at the same time, is bare of their flaws.
Specifically, it resums large logarithms by the
Dokshitzer-Gribov-Lipatov-Altarelli-Parisi (DGLAP) evolution of
nonperturbative fragmentation functions (FFs), guarantees the universality of
the latter as in the ZM-VFN scheme, and simultaneously retains the
mass-dependent terms of the FFN scheme without additional theoretical
assumptions.
It was elaborated at next-to-leading order (NLO) for photoproduction \cite{KS}
and hadroproduction \cite{KKSS} of  charmed hadrons as well as for their
production by $e^+e^-$ annihilation \cite{Kneesch:2007ey}.
It was also applied to obtain predictions for $B$-meson hadroproduction
\cite{KKSS3}, which could be compared with recent CDF data
\cite{Acosta:2004yw}.
An earlier implementation of such an interpolating scheme is the so-called
fixed-order-next-to-leading-logarithm (FONLL) approach, in which the
conventional cross section in the FFN scheme is linearly combined, with the
help of a $p_T$-dependent weight function, with a suitably modified cross
section in the ZM-VFN scheme implemented with perturbative FFs \cite{CGN}.

In Ref.~\cite{KKSS3}, nonperturbative FFs for the transitions $a\to B$, where
$a$ is any parton, including $b$ and $\bar{b}$ quarks, were extracted at NLO in
the $\overline{\rm MS}$ factorization scheme with $n_f = 5$ flavors from the
scaled-energy ($x$) distributions $d\sigma/dx$ of $e^+e^- \to B + X$ measured
by the ALEPH \cite{Heister:2001jg} and OPAL \cite{Abbiendi:2002vt}
Collaborations at the CERN LEP1 collider and by the SLD Collaboration
\cite{Abe:2002iq} at the SLAC SLC collider.
As explained in Ref.~\cite{KKSS3}, these FFs may be consistently used in our
GM-VFN framework.
Working at NLO in the GM-VFN scheme with these $B$-meson FFs, we found
excellent agreement with recent CDF measurements of $d\sigma/dp_T$ for
$p\bar{p}\to B+X$ \cite{Acosta:2004yw}, especially in the upper $p_T$ range,
$p_T \gsim 10$ GeV \cite{KKSS3}.

The content of this paper is as follows. In Sec.\ 2, we summarize our 
input choices of PDFs and $B$-meson FFs. In 
Sec.\ 3, we compare the predictions of the GM-VFN scheme with the CMS data 
from the recent LHC run at $\sqrt{S} = 7$ TeV 
\cite{Khachatryan:2011mk,Chatrchyan:2011pw,Chatrchyan:2011vh}. 
Our conclusions are given in Sec.\ 4.

%**********************************************************************

\boldmath
\section{Input PDFs and $B$-meson FFs}
\unboldmath

As PDFs for the proton, we choose one of the most recent parametrizations 
of the CTEQ Collaboration, set CTEQ6.6M \cite{CTEQ6.6}, which provides 
an improvement over the earlier version CTEQ6.5M. Both sets were obtained in 
the framework of a general-mass scheme using the input values $m_c = 1.3$ 
GeV, $m_b = 4.5$ GeV, and $\alpha_s(m_Z)=0.118$. In both set, 
the $b$-quark PDF has its starting scale at $\mu_0 = m_b$. 

The nonperturbative FFs describing the transition of the $b$ and $\bar b$ 
quarks into a $B$ meson can be obtained only from experiment. In our 
earlier work on inclusive $B$-meson production at the 
Tevatron \cite{KKSS3}, we constructed such FFs using as input recent precise 
measurements of the cross section of inclusive $B$-meson production 
in $e^+e^-$ annihilation obtained by the ALEPH \cite{Heister:2001jg}, 
OPAL \cite{Abbiendi:2002vt}, and SLD \cite{Abe:2002iq} 
Collaborations.\footnote{
  Recently, similar data became available also from the DELPHI 
  Collaboration \cite{DELPHI:2011ep}.
} These data were taken 
on the $Z$-boson resonance, so that finite-$m_b$ effects, being of relative 
order $m_b^2/m_Z^2$, are strongly suppressed, which means that we are 
in the asymptotic regime where the GM-VFN scheme is equivalent to the ZM-VFN
scheme. 
The combined fit to the three data sets was performed using the NLO value 
$\Lambda_{\overline{MS}}^{(5)} = 227$ MeV corresponding to 
$\alpha_s^{(5)}(m_Z) = 0.1181$, values adopted from 
Ref.~\cite{CTEQ6.6}. The renormalization and factorization scales were 
chosen to be $\mu_R = \mu_F = m_Z$. In accordance with the chosen 
PDFs, the starting scale of the $b \rightarrow B$ FF 
was taken to be $\mu_0 = m_b$, while the $g,q \to B$ FFs, where $q$ 
denotes the light quarks including the charm quark, were taken to vanish 
at $\mu_F = \mu_0$. 

For fitting the data, we actually employed two different 
parametrizations for the $b \to B$ FF at $\mu_0 = m_b$, namely the
Peterson ansatz \cite {Peterson} and the simple power ansatz \cite{power}. 
It turned out that the Peterson ansatz led to a very poor fit. Therefore, 
we shall use in this work only the FFs obtained with the power ansatz, 
whose parameters at the starting scale are listed in Table 1 of 
Ref.~\cite{KKSS3}. A comparison of the fit performed using this ansatz with
the three input data sets may be found in Fig.\ 1 of that reference.

We note that the data from OPAL and SLD included all $B$-hadron 
final states, in particular those with $\Lambda_b$ hadrons, while, in 
the ALEPH analysis, only final states with identified $B^{\pm}$ and 
$B^0$ mesons were taken into account. Our fit was based on the 
assumption that the FFs of all $b$ hadrons had the same shape. The  
branching fraction of $b \rightarrow B^+$ was taken equal to that 
of $b \rightarrow B^0$ and fixed to 0.397. In our calculations for 
$B_s$-meson production to be presented below, we shall use the same 
FFs and rescale them by the factor $0.113 / 0.401$, which uses the
up-to-date values for the $b \rightarrow B^+$ and $b \rightarrow B_s$
branching fractions quoted by the Particle Data Group
\cite{Nakamura:2010zzi}.

We should emphasize that, in the analysis of the available $e^+e^-$ 
annihilation data, the charged and neutral $B$ mesons were not
separated. Furthermore, the charged states $B^+$ and $B^-$ could not be 
distinguished. The FFs obtained in Ref.~\cite{KKSS3} are, therefore, valid 
for the average of $B^+$ and $B^-$ and, similarly, for the average of 
$B^0$ and $\overline{B^0}$.

The factorization scales related to the initial- and final-state 
singularities entering the PDFs and FFs, respectively, can in principle 
be chosen independently. We checked, however, that when estimating 
theoretical error bands by varying these scales by factors of 2 up and 
down, the extreme values are indeed obtained when the initial- and 
final-state factorization scales are identified.
Our default choice of renormalization and factorization scales is 
$\mu_R = \mu_F = m_T = \sqrt{p_T^2+m_b^2}$. Theoretical uncertainties 
will be estimated by setting $\mu_R = \xi_R m_T$ and $\mu_F = \xi_F m_T$, 
and varying $\xi_R$ and $\xi_F$ about their default values $\xi_R = 
\xi_F = 1$ by factors of 2 up and down, restricting the ratio to the 
range $1/2 \leq \xi_R/\xi_F \leq 2$.

%**********************************************************************

\boldmath
\section{Theoretical Predictions for $pp \to B+X$ and Comparisons with 
CMS Data}
\unboldmath

%----------------------------------
\begin{figure}[b!] 
\begin{center}
\includegraphics[scale=0.73]{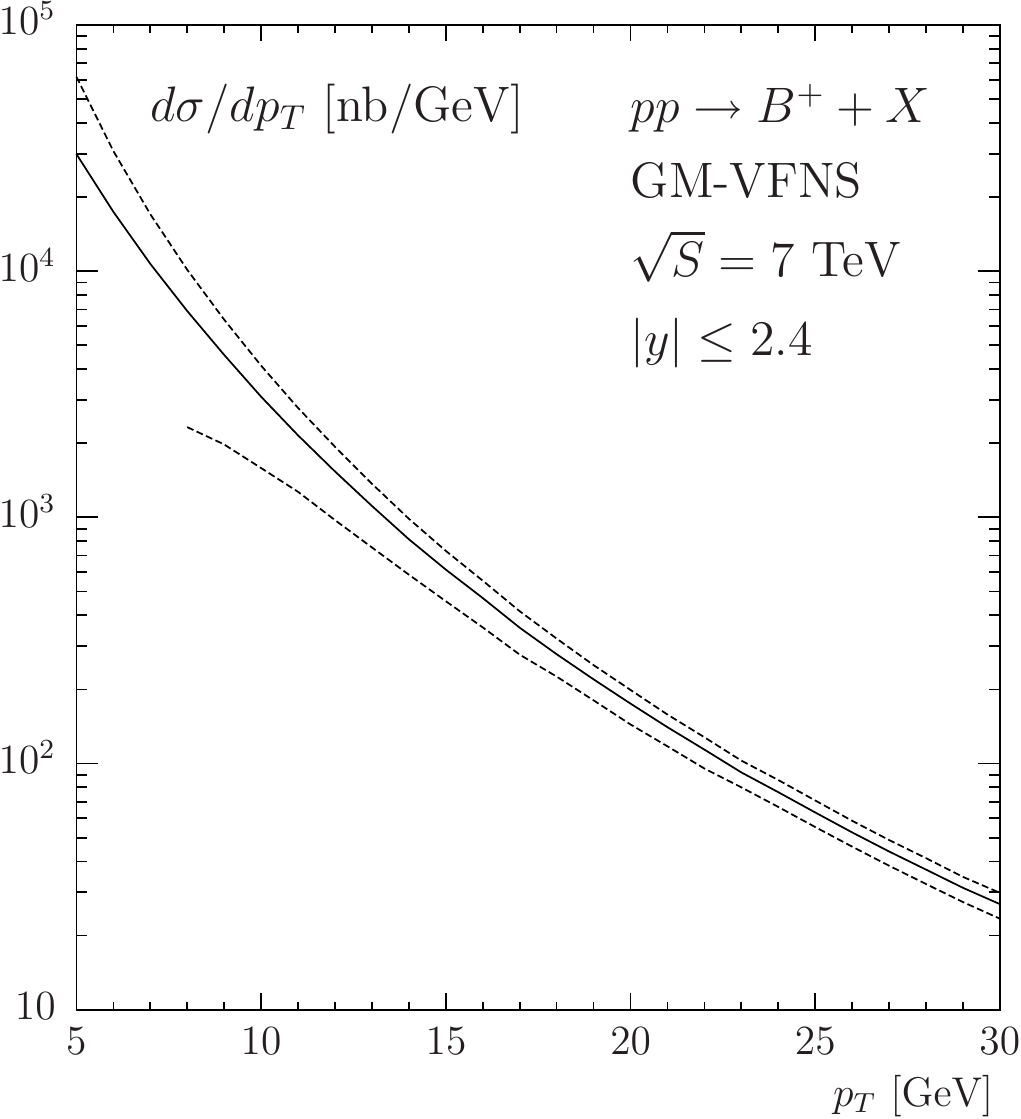}
\includegraphics[scale=0.73]{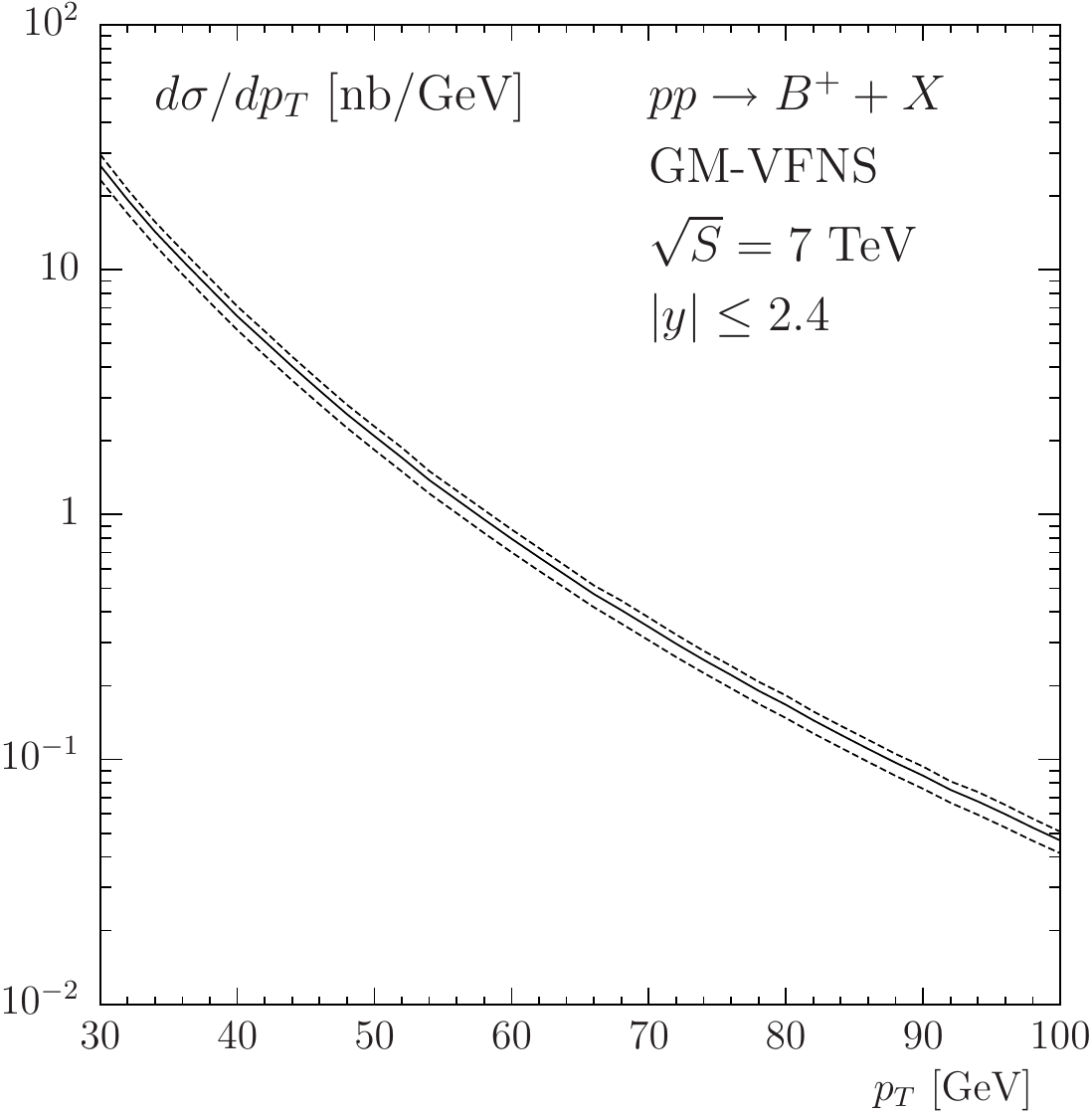}
\end{center}
\caption{
  $d\sigma/dp_T$ [nb/GeV] for $pp \rightarrow B^++X$ at $\sqrt{S} = 
  7$ TeV in the GM-VFNS. For clarity, we split the $p_T$ range into 
  a lower part ($p_T$ below 30 GeV, left panel) and an upper part 
  ($p_T$ above 30 GeV, right panel). The central values (solid 
  lines) correspond to the default choice of scale parameters, $\xi_R 
  = \xi_F = 1$. An error band (dashed lines) is obtained from variations 
  of the renormalization and factorization scales by factors of 2 up and 
  down. The upper end of the error band is reached for $\xi_R = 1$ and 
  $\xi_F = 2$ at $p_T < 21$ GeV and for $\xi_R = 0.5$ and $\xi_F = 1$ 
  at $p_T > 21$ GeV, the lower error end is reached for $\xi_R = 1$ and 
  $\xi_F = 0.5$ at $p_T < 25$ GeV and $\xi_R = 2$ and $\xi_F = 1$ at 
  $p_T > 25$ GeV.
}
\label{fig1}
\end{figure}
%----------------------------------

To obtain an overview of the $p_T$ dependence of $d\sigma/dp_T$, we 
first show results for this observable, integrated over $|y| \leq 2.4$, 
for the case of $B^+$ production in the GM-VFN scheme as described above.
This differential cross 
section is shown in Fig.\ \ref{fig1} (left) for $p_T$ values between 5 and 
30 GeV and in Fig.\ \ref{fig1} (right) for larger $p_T$ values, 
up to 100 GeV, where we expect data to come in the near future when 
the LHC experiments are accumulating more statistics.

In the $p_T$ range between 5 and 30 GeV, the cross section falls off by 
three orders of magnitude. This is essentially due to the behavior 
of the PDFs as a function of the scaling variable $x$ and less so from 
the behavior of the partonic cross sections. Towards low $p_T$ values, both 
the upper edge of the error band and the cross section for the default 
choice of scales rise steadily with decreasing $p_T$ value, down to $p_T=5$ 
GeV. This is caused by the scale dependence of the $b$-quark PDF and the FFs.
With our choice of scales, they fade out 
and quench the cross section, leading to a turn-over of the $p_T$ 
distributions only at $p_T=0$ and not already at some finite $p_T$ value. 
The lower edge of the error band is obtained for $\xi_F=0.5$. Here, 
both the $b$-quark PDF and the FFs vanish at $p_T \approx 8$ GeV, 
corresponding to $\mu_F = m_b = 4.5$ GeV. The line representing the 
lower edge of the error band therefore stops at this point.

%----------------------------------
\begin{figure}[b!] 
\begin{center}
\includegraphics[scale=0.73]{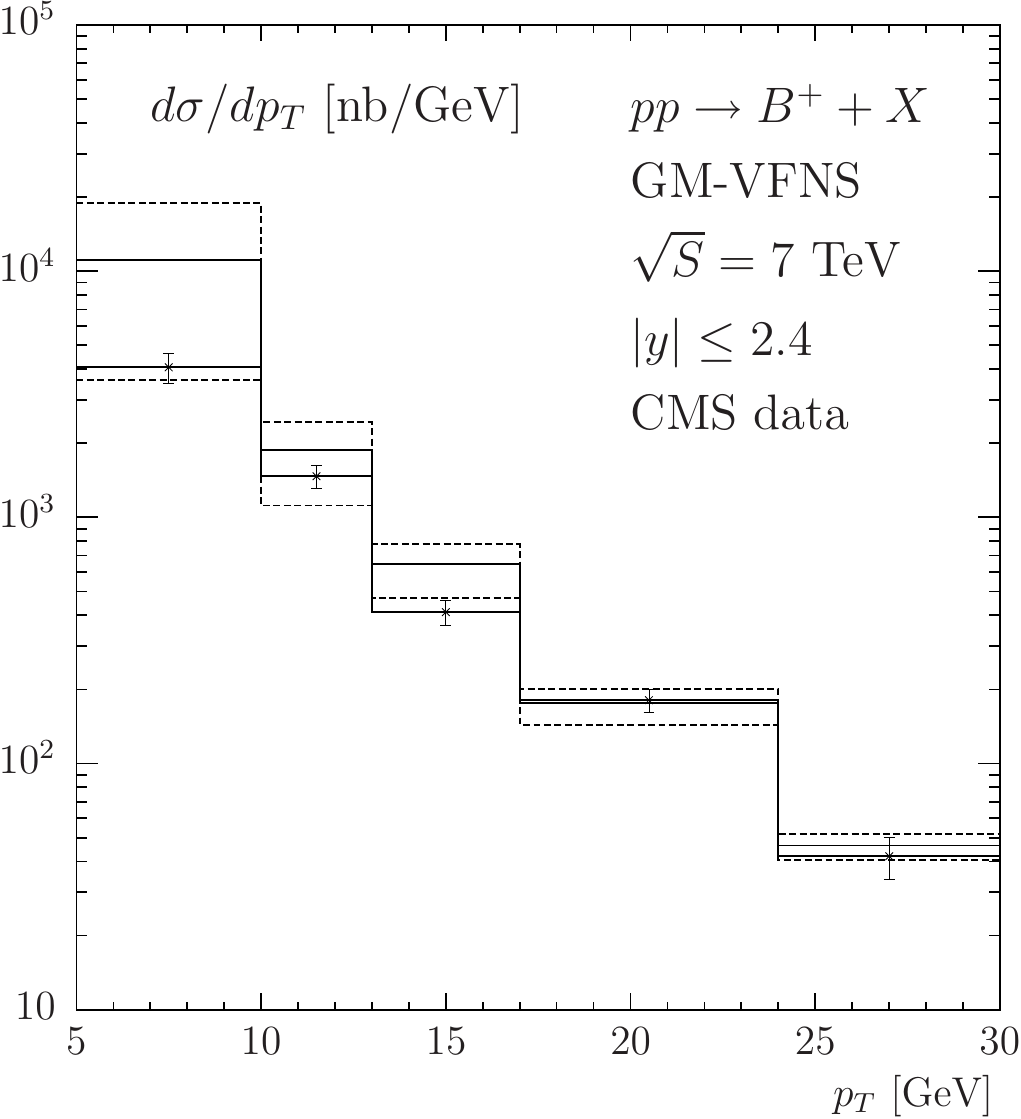}
\includegraphics[scale=0.73]{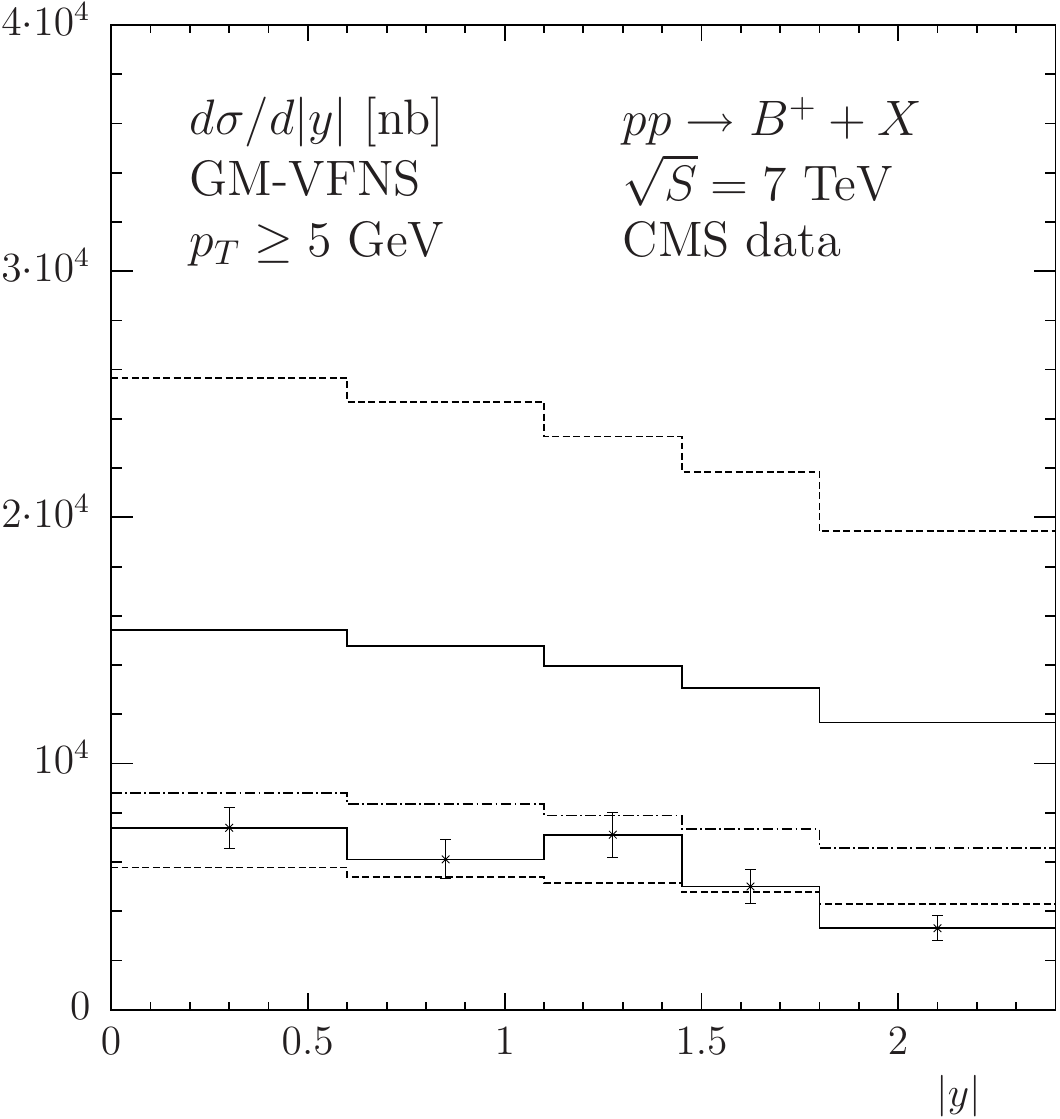}
\end{center}
\caption{
  $d\sigma/dp_T$ [nb/GeV] (left panel) and $d\sigma/d|y|$ [nb] (right 
  panel) for $pp \to B^++X$ at NLO in the GM-VFN scheme compared with the 
  CMS data \cite{Khachatryan:2011mk}. The central values (solid lines)
  correspond to the choice 
  $\xi_R = \xi_F = 1$. We also show the prediction for $d\sigma/d|y|$ obtained
  with the choice $\xi_R = 1$ and $\xi_F = 0.7$ (dash-dotted line). The 
  error bands (dashed lines) are obtained by varying $\xi_R$ and $\xi_F$ by
  factors of 2 up 
  and down (maximum: $\xi_R = 1$, $\xi_F = 2$; minimum: $\xi_R=1$, $\xi_F=0.5$).
}
\label{fig2}
\end{figure}
%----------------------------------

The CMS Collaboration measured the differential cross section
$d\sigma/dp_T$ for the production of $B^+$ mesons \cite{Khachatryan:2011mk}
(actually the average of $B^+$ and $B^-$ mesons), 
integrated over the $y$ range $|y| \leq 2.4$, as a function of $p_T$.
The measurement 
covered the $p_T$ range from 5~GeV to 30~GeV with five bins. In 
addition, the differential cross section $d\sigma/d|y|$, integrated over the
considered $p_T$ range, was
given for six $|y|$ bins. In Ref.~\cite{Chatrchyan:2011pw}, the 
results of the measurement of $B^0$-meson production (again for the average 
of the charge-conjugate states $B^0$ and $\overline{B^0}$) were 
presented. They comprise the differential cross section $d\sigma/dp_T$,
integrated over the $y$ range $|y| \leq 2.2$, in five $p_T$ bins between
$p_T = 5$ GeV and $p_T = 40$ GeV and $d\sigma/d|y|$, integrated over the
considered $p_T$ range, in five 
$|y|$ bins. Since, in this second 
analysis, a larger luminosity was already available, the $B^0$ data 
extend to larger $p_T$ values.

%----------------------------------
\begin{figure}[t!] 
\begin{center}
\includegraphics[scale=0.73]{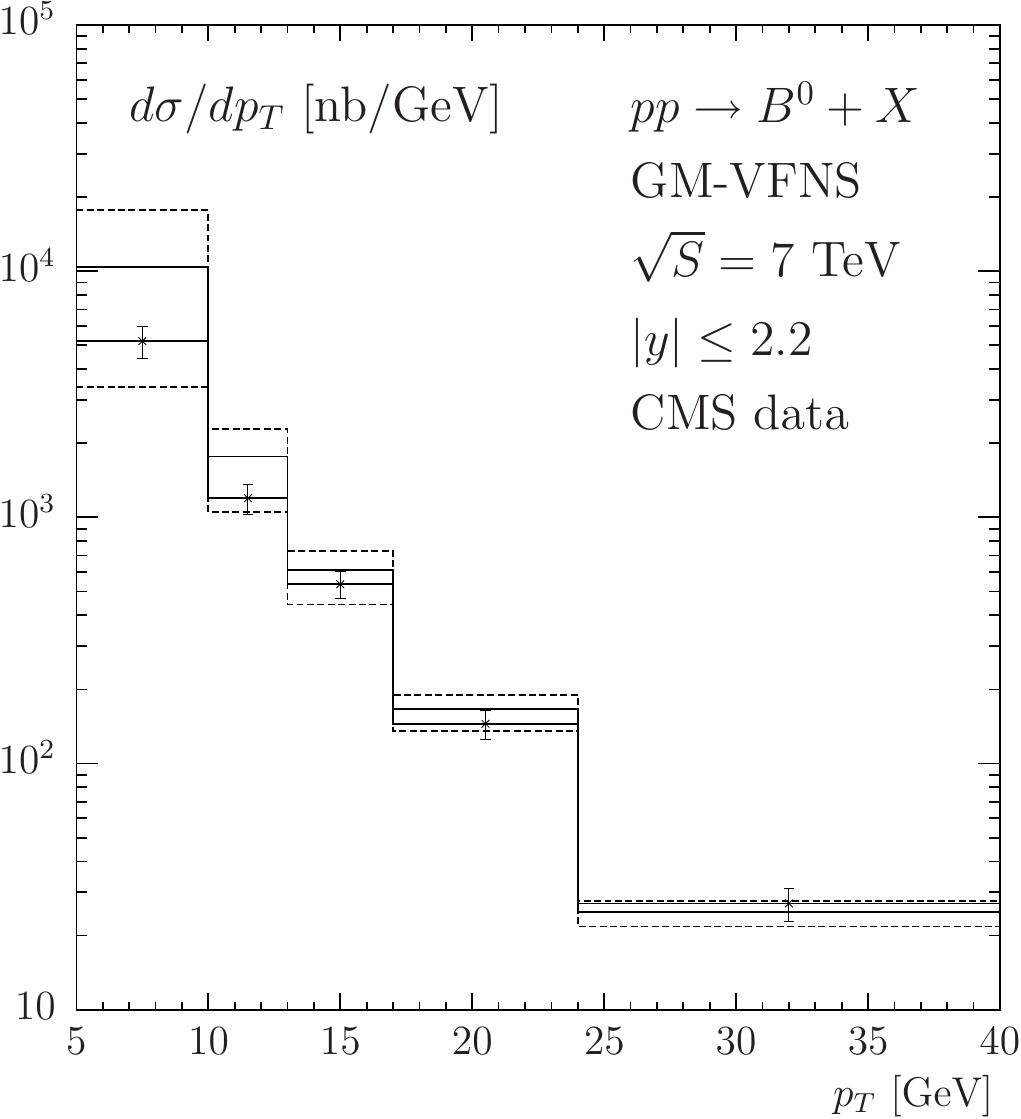}
\includegraphics[scale=0.73]{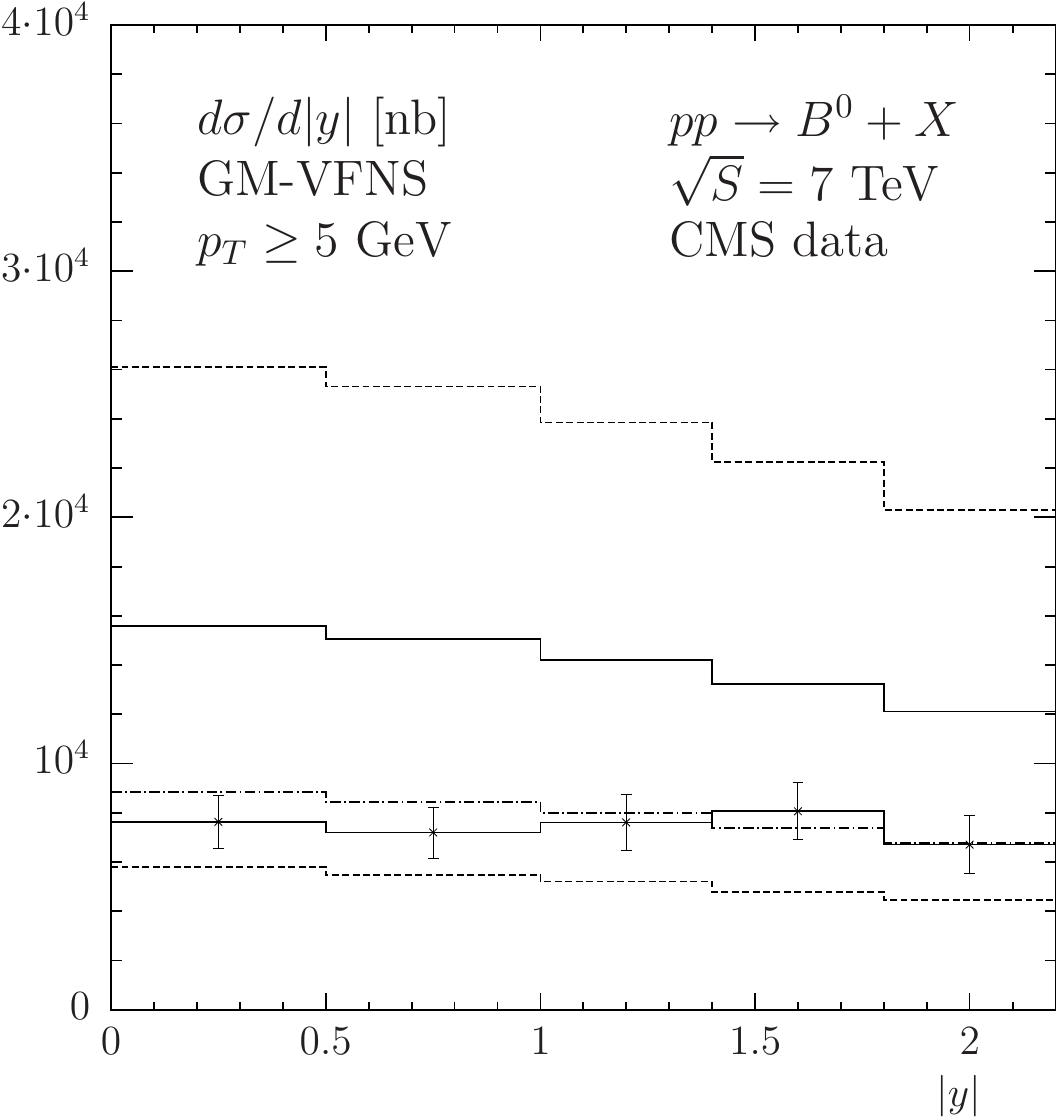}
\end{center}
\caption{
  $d\sigma/dp_T$ [nb/GeV] (left panel) and $d\sigma/d|y|$ [nb] (right 
  panel) for $pp \to B^0+X$ at NLO in the GM-VFN scheme compared with the 
  CMS data \cite{Chatrchyan:2011pw}. The central values (solid lines)
  correspond to the choice $\xi_R = \xi_F = 1$.
  We also show the prediction for $d\sigma/d|y|$ obtained
  with the choice $\xi_R = 1$ and $\xi_F = 0.7$ (dash-dotted line). The 
  error bands (dashed lines) are obtained by varying $\xi_R$ and $\xi_F$ by
  factors of 2 up 
  and down (maximum: $\xi_R = 1$, $\xi_F = 2$; minimum: $\xi_R=1$, $\xi_F=0.5$).
}
\label{fig3}
\end{figure}
%----------------------------------

In order to facilitate the comparisons with the CMS measurements
\cite{Khachatryan:2011mk,Chatrchyan:2011pw}, we integrate over the bins using
the same binnings.
The $p_T$ bins for $B^+$- and $B^0$-meson production are the same, except
for the largest one.
Our results are shown in Figs.\ 
\ref{fig2} and \ref{fig3}, where they are compared with the experimental 
data. The errors of the experimental data points are obtained 
from Ref.~\cite{Chatrchyan:2011pw} by adding in quadrature the statistic and
systematic errors quoted there.
The differences between the predictions in Figs.\ \ref{fig2} and \ref{fig3}
are entirely due to the different bin choices, the FFs being the same in both
cases.

We determine the error band from variations of the scale parameters 
by factors of 2 as described above, except that the minimum of the theoretical 
prediction is obtained with the additional prescription that the FFs are 
frozen when $\mu_F$ falls below the starting scale $\mu_0 = m_b$.
Otherwise the cross section would 
become zero for $\xi_F=0.5$ in a large part of the first $p_T$ bin, so that
the lower edge of the error band would become meaningless. As is seen in
Figs.\ \ref{fig2} and \ref{fig3}, the data lie inside the error bands. 
In the case of $B^+$ ($B^0$) mesons, the default predictions appreciably
overshoot the CMS data in the first three (two) $p_T$ bins, while they are
very close to the CMS data in the residual $p_T$ bins.
The default values 
of the predicted cross sections are a factor of approximately 2 (1.5) 
larger than the experimental central values in the lowest (next-to-lowest) 
$p_T$ bins. This is caused by the fact that, with our choice of scales, 
large contributions coming from initial-state $b$ quarks are present for all 
finite values of $p_T$. If one changes the factorization scale to a lower 
value, for example by setting $\xi_R=1$ and $\xi_F = 0.7$, the $b$-quark PDF vanishes 
at $p_T = 4.6$ GeV. Furthermore, with our prescription, the PDFs and the 
FFs are frozen at the values they reach at $\mu_F = m_b$ when $p_T$ falls 
below $p_T = 4.6$ GeV. For this special choice of factorization 
scales, we obtain the cross section values given for the $B^0$-meson case in
the column 
labeled $\xi_R=1$, $\xi_F = 0.7$ of Tab.\ 1. For comparison, we present the experimental 
results in the second column of this table and the default-scale results of
Fig.\ \ref{fig3} (left) in the third one.
We see that the theoretical values of the cross sections in the five 
$p_T$ bins agree with the experimental values quite well, within the errors.
The total $B^0$-meson production cross section determined by CMS in the
considered kinematic range is
$\sigma_{\rm tot} = 33.2 \pm 4.3~\mu$b. For the 
default choice of scales $\xi_R = \xi_F = 1$, we find $\sigma_{\rm tot} = 
61.7\,\mu$b, while the result for $\xi_R=1$ and $\xi_F = 0.7$ is 35.0~$\mu$b, in 
very good agreement with the data.
A similar comparison may be performed 
for $pp \to B^{+}X$, with similar conclusions, as can be inferred 
from Fig.\ \ref{fig2} (right panel), where we show the corresponding 
results for $d\sigma/d|y|$. The theoretical predictions are almost identical, 
since the FFs for $b \rightarrow B^+$ and $b \rightarrow B^0$ are taken to be
the same and there is only a tiny difference due to the different upper ends
of the $p_T$ ranges.

%----------------------------------
\begin{table}[t!]
\begin{center}
\label{tab:mom}
\bigskip
\begin{tabular}{ccccc}
\hline\hline
$p_T$ (in GeV) 
      & Data \cite{Chatrchyan:2011pw} 
                     & $\xi_R=\xi_F=1$ 
                             & $\xi_R=1$, $\xi_F=0.7$ 
                                    & $\xi_a=0.2$ \\
\hline
 5--10 & $5200\pm770$ & 10356 & 5578 & 6327 \\
10--13 & $1196\pm168$ &  1769 & 1265 & 1016 \\
13--17 & $ 535\pm 68$ &   610 &  481 &  401 \\
17--24 & $ 145\pm 20$ &   166 &  141 &  124 \\
24--40 & $  27\pm  4$ &    25 &   22 &   21 \\
\hline\hline
\end{tabular}
\caption{
  Predictions for the differential cross section $d\sigma/dp_T$ [nb/GeV] of
  $B^0$-meson production with different renormalization and factorization
  scales compared with the
  CMS data \cite{Chatrchyan:2011pw}, for which the statistical and 
  systematic errors are added in quadrature. The values presented in the
  second and third columns are also displayed in Fig.~\ref{fig3} (left). 
}
\end{center}
\end{table}
%----------------------------------

As explained above, massless contributions, in particular the ones 
due to incoming $b$ quarks, dominate the total cross section
towards low $p_T$ values. 
These contributions lead to an increase of $d\sigma/dp_T$ in the limit $p_T 
\rightarrow 0$ because the heavy-quark PDFs carry resummed 
logarithms, which are not fully cancelled by the subtraction terms in 
the GM-VFN approach, which are implemented at NLO, i.e.\ at 
fixed order only. This increase can be tamed by imposing the kinematic 
cut $\hat{s} > 4m_b^2$ for the partonic center-of-mass energy $\hat{s}$ 
also for the massless contributions. Furthermore, a judicious choice 
of the factorization scale, e.g.\ 
\begin{equation}
\mu_F = \sqrt{m_b^2 + \xi_a p_T^2},
\label{eq:scale}
\end{equation}
with a parameter $\xi_a < 1$, can boost the transition
$\mu_F \to \mu_0 = m_b$ for $p_T \to 0$. This prescription creates a turn-over 
of the $p_T$ distribution towards low $p_T$ values and also allows us to obtain
a reasonable description of the CDF data 
\cite{Acosta:2004yw}, which were 
taken at lower $p_T$ values. The CMS data start at $p_T = 5$ GeV, and a 
turn-over is not visible in $d\sigma/dp_T$. However, the ansatz of
Eq.~(\ref{eq:scale}) leads to a reduction of the $p_T$ distribution for small
$p_T$ values, i.e.\ to a significant change of $d\sigma/dp_T$ in the first two
$p_T$ bins. 
The cross section values obtained for $B^0$ mesons using the scale choice of
Eq.~(\ref{eq:scale}) with $\xi_a = 0.2$ are presented in the last column of
Tab.\ 1. We 
find that this approach leads to a better description of the CMS data, which
is, however, not as good as for the scale choice $\xi_F = 0.7$ (fourth column
of Tab.\ 1).

%----------------------------------
\begin{figure}[t!] 
\begin{center}
\includegraphics[scale=0.73]{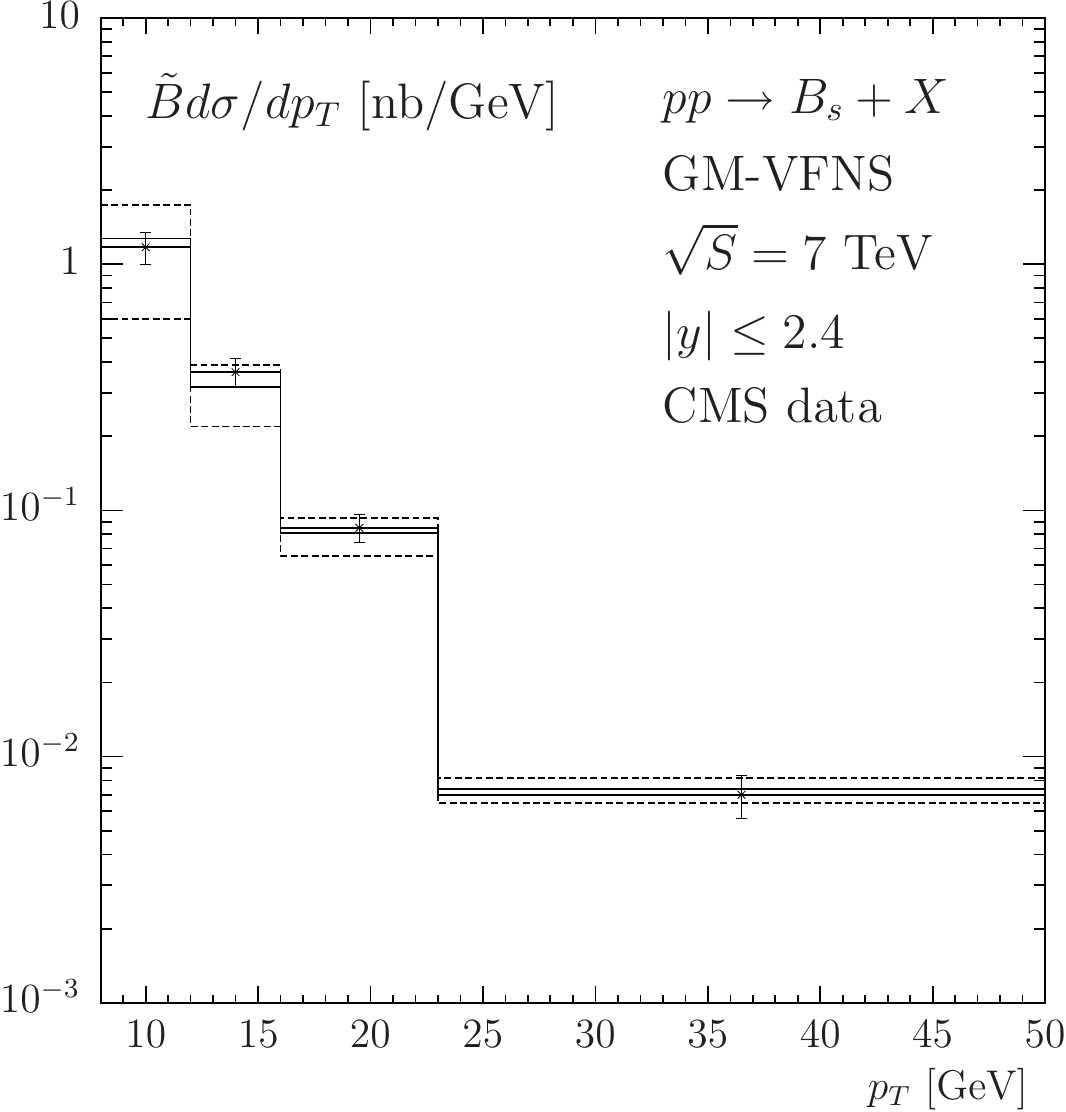}
\includegraphics[scale=0.73]{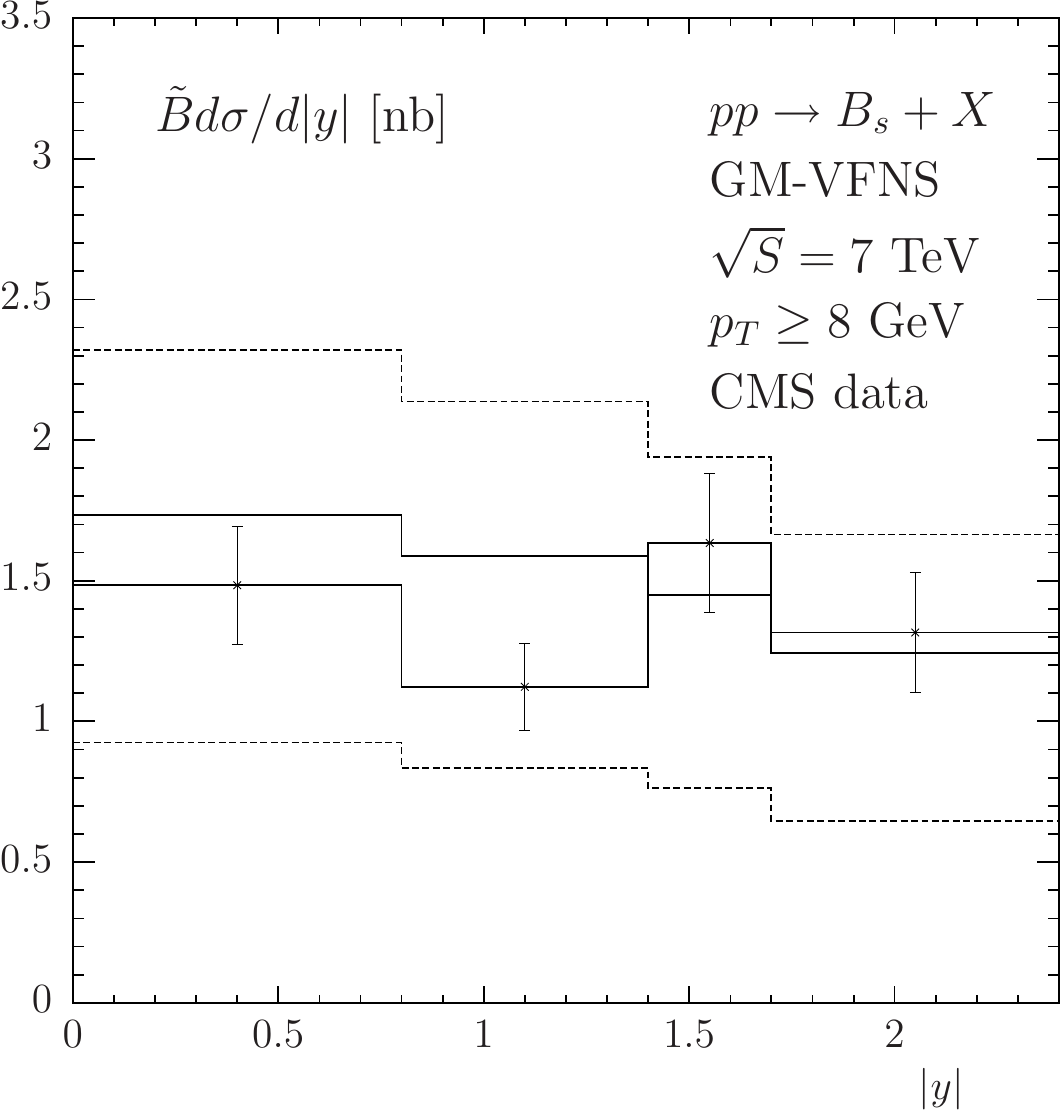}
\end{center}
\caption{
  $\tilde{B}d\sigma/dp_T$ [nb/GeV] (left panel) and $\tilde{B}d\sigma/d|y|$
  [nb] (right 
  panel) for $pp \to B_s+X$ at NLO in the GM-VFN scheme compared with the 
  CMS data \cite{Chatrchyan:2011vh}.
  The branching fraction of the decay $B_s \rightarrow J/\psi \phi$ is assumed
  to be $\tilde{B} = 1.3 \times 10^{-3}$ \cite{Nakamura:2010zzi}. 
  The central values (solid lines)
  correspond to the choice $\xi_R = \xi_F = 1$. The 
  error bands (dashed lines) are obtained by varying $\xi_R$ and $\xi_F$ by
  factors of 2 up 
  and down (maximum: $\xi_R = 0.5$, $\xi_F = 1$; minimum: $\xi_R=1$,
  $\xi_F=0.5$).
}
\label{fig4}
\end{figure}
%----------------------------------

As a side remark, we note that the behavior towards small $p_T$ values is 
not due to a shift in the average $B$-meson to $b$-quark momentum fraction.
This may be observed by calculating the quantity
\begin{equation}
\langle z \rangle (p_T) = 
\frac{\int dz\, z d\sigma (p_T)}{\int dz\, d\sigma (p_T)},
\nonumber
\end{equation}
where $z$ is the scaling variable of the FFs and it is understood that the
integration is also done over the rapidity interval $|y| \leq 2.4$ relevant
for the CMS measurement \cite{Khachatryan:2011mk}.
We find 
a rather weak dependence on $p_T$. In fact, $\langle z \rangle$ decreases from 
0.770 at $p_T = 5$ GeV to 0.749 at $p_T = 30$ GeV, which means that, 
in our applications, the $b\to B$ FF is always probed around its maximum
(see Ref.\ \cite{KKSS3}).

We now discuss the $|y|$ distributions $d\sigma/d|y|$ of $B^+$ and $B^0$
production shown in the right panels of Figs.~\ref{fig2} and \ref{fig3},
respectively.
The bulk of these cross sections comes from the lowest $p_T$ bin, where the
theoretical uncertainties are largest, as is evident from Tab.\ 1.
However, it is interesting to find out how much the shapes of these
differential cross sections depend on the various scale choices.
In order to get some idea about this, we include in the right panels of
Figs.~\ref{fig2} and \ref{fig3} as dot-dashed histograms also the predictions
evaluated using the scale choice $\xi_R=1$ and $\xi_F = 0.7$, as in the
fourth column in Tab.\ 1.
They agree fairly well with the CMS data, while the default predictions
($\xi_R = \xi_F = 1$), shown as solid histograms, significantly overshoot the
CMS data as expected, but their shapes are still reasonable. 

Finally, in Fig.\ \ref{fig4}, we present our predictions for the production of
$B_s$ mesons and compare them with the experimental data published by
the CMS Collaboration in Ref.\ 
\cite{Chatrchyan:2011vh}. $d\sigma/dp_T$ was measured in four $p_T$ bins 
between $p_T = 8$ and 50 GeV and integrated over $|y| \leq 2.4$, and 
$d\sigma/d|y|$ was measured in four $|y|$ bins spanning this $|y|$ range and
integrated over the full $p_T$ range considered.
Both the experimental data and our theoretical predictions refer to the product
of cross section times branching fraction $\tilde{B}$ for $B_s \to J/\psi \phi$,
for which we adopt the value $1.3 \times 10^{-3}$ from
Ref.~\cite{Nakamura:2010zzi}. 
In this case, we find better agreement between theory and experiment over the 
full $p_T$ range, probably due to the fact that very low values of $p_T$, with  
$p_T < 8$~GeV, are excluded from this analysis. The total cross section 
times branching fraction measured by CMS for 8 GeV ${}\leq p_T 
\leq 50$~GeV and $|y| \leq 2.4$ is $6.9 \pm 0.8$~nb, while our calculation 
yields 7.2~nb. 

%**********************************************************************

\section{Conclusions}

In summary, we applied the GM-VFN
scheme to obtain NLO predictions for the production of $B$ mesons in $pp$ 
collisions at the LHC. The comparison with experimental data from 
the CMS Collaboration at $\sqrt{S} = 7$ TeV generally shows good 
agreement between theory and experiment, in particular at large $p_T$ values. 
The agreement is particularly good for the case of $B_s$-meson 
production, where data are restricted to $p_T$ values above 8 GeV. 
At low $p_T$ values, we observe large scale uncertainties. 

Future data collection at the LHC will allow us to extend the comparisons 
with theoretical predictions to much wider $p_T$ ranges.
If also the systematic uncertainties can be further reduced, 
we may expect that $B$-meson production will play an increasingly 
important role in constraining size and shape of both PDFs and FFs.

%**********************************************************************

%%%%%%%%%%%%%%%%%%%%%%%%%%%%%%%%%%%%%%%%%%%%%%%%%%%%%%%%%%%%%%%%%%%%%%%%%%%%%
%%%%%%%%%%%%%%%%%%%%%%%%%%%%%%%%%%%%%%%%%%%%%%%%%%%%%%%%%%%%%%%%%%%%%%%%%%%%%
%%%%%%%%%%%%%%%%%%%%%%%%%%%%%%%%%%%%%%%%%%%%%%%%%%%%%%%%%%%%%%%%%%%%%%%%%%%%%

\end{document}